\documentclass[twocolumn,showpacs,aps]{revtex4}
\usepackage[dvips]{graphicx}
\usepackage{bm}
\usepackage{mathrsfs}
\usepackage{amssymb}
\begin{document}
\date{\today}

\title{Algebraic construction of spherical harmonics}
\author{Naohisa Ogawa
\footnote{ogawanao@hus.ac.jp}}
\affiliation{Hokkaido University of Sciences, Sapporo 006-8585 Japan}

\begin{abstract}
The angular wave functions for a hydrogen atom are well known to be spherical harmonics, and are obtained as the solutions of a partial differential 
equation. However, the differential operator is given by the Casimir operator 
of the $SU(2)$ algebra and its eigenvalue $l(l+1) \hbar^2$, where $l$ is non-negative integer, is easily obtained
 by an algebraic method. Therefore the shape of the wave function may also be
 obtained by extending the algebraic method. In this paper, we describe the 
 method and show that wave functions with different quantum numbers are 
 connected by a rotational group in the cases of $l=0$, 1 and 2.
\end{abstract}
\maketitle

\section{Introduction- Representation of Angular moment}

Spherical harmonics (hereafter abbreviated to SHs,)
$Y_{l,m}(\theta,\phi)$ are usually obtained by solving the following partial
differential equation using the Laplacian on $S_2$ 
\cite{Schiff}-\cite{Griffiths}:
\begin{equation}
[-\frac{1}{\sin \theta}\frac{\partial}{\partial \theta}
(\sin \theta \frac{\partial}{\partial \theta}) - \frac{1}{\sin ^2 \theta}
\frac{\partial^2 }{\partial \phi^2}] Y(\theta,\phi) = \lambda ~ Y(\theta,\phi),\label{eq:Laplacian}
\end{equation}
where $\lambda$ is the real eigenvalue, which is later shown to be $l(l+1)$.
However, there is another method of solving this equation by using algebra, 
that does not depend on the choice of coordinates.
In this algebraic method the eigenvalues are easily obtained 
but eigenfunctions (angular wave functions: SHs) cannot be calculated.
The purpose of this manuscript is to show how to obtain SHs 
by algebraic method.

In this section, we briefly sketch the conventional algebraic method 
to provide a self-contained explanation \cite{Schiff}-\cite{Chisholm}.
Then using the ideas, tools, and notations expressed here, 
we will show how to obtain the wave function.
Hereafter, we utilize the natural unit $\hbar=1$ for simplicity.

Note that the differential operator in the l.h.s. of (\ref{eq:Laplacian})
is given by
\begin{equation}
\hat{\vec{L}}^2 = \hat{L}_x^2 + \hat{L}_y^2 +\hat{L}_z^2,
\end{equation}
with
\begin{eqnarray}
\hat{L}_x &=& -i(y \frac{\partial}{\partial z} - z \frac{\partial}{\partial y})\nonumber\\
&=& i (\sin \phi \frac{\partial}{\partial \theta} + \cot \theta \cos \phi \frac{\partial}{\partial \phi}),\label{eq:Lx}\\
\hat{L}_y &=& -i(z \frac{\partial}{\partial x} - x \frac{\partial}{\partial z})\nonumber\\
&=& i (-\cos \phi \frac{\partial}{\partial \theta} + \cot \theta \sin \phi \frac{\partial}{\partial \phi}),\label{eq:Ly}\\
\hat{L}_z &=& -i(x \frac{\partial}{\partial y} - y \frac{\partial}{\partial x})=-i \frac{\partial}{\partial \phi},\label{eq:Lz}
\end{eqnarray}

where $\hat{L}_i~ (i=1,2,3)$ satisfies the $SU(2)$ algebra:
\begin{equation}
[\hat{L}_i, \hat{L}_j] = i \epsilon_{ijk} \hat{L}_k.
\end{equation}
$\epsilon_{ijk}$ is the absolute antisymmetric unit tensor with 
$\epsilon_{123}=\epsilon_{231}=\epsilon_{312}=1, 
~\epsilon_{213}=\epsilon_{132}=\epsilon_{321}=-1$, and other terms
 equal to 0.
We find that $\hat{\vec{L}}^2$ is a Casimir operator:
\begin{equation}
[\hat{\vec{L}}^2, \hat{L}_i] =0 ~(i=1,2,3).
\end{equation}
We can diagonalize one of the three angular momentum operators.
Usually we select $\hat{L}_z$ to be diagonal.

Furthermore, by using the notation
\begin{equation}
\hat{L}_{\pm} = \hat{L}_x \pm i \hat{L}_y,
\end{equation}
we obtain
\begin{equation}
[\hat{L}_z, \hat{L}_{\pm}]= \pm \hat{L}_{\pm}. \label{eq:updown}
\end{equation}

Let us consider the following eigenvalue equation of $\hat{L}_z$:

\begin{equation}
\hat{L}_z |m> = m |m>,
\end{equation}
where the ket vector $|m>$ denotes the eigenstate with eigenvalue $m$ 
\cite{Dirac}.
For consistency with (\ref{eq:updown}), we obtain
\begin{equation}
\hat{L}_z (\hat{L}_{\pm}|m>) = (m \pm 1) (\hat{L}_{\pm}|m>).
\end{equation}
This means we have a new ket with an eigenvalue that differs by $\pm 1$.

\begin{equation}
|m\pm1> \sim \hat{L}_{\pm} | m>,
\end{equation}
where $\sim$ denotes the ambiguity of the constant coefficient, which 
should be determined from the normalization conditions except total phase.
We consider the maximum state $m_{max} =l$ that satisfies
\begin{equation}
\hat{L}_+ |l> = 0
\end{equation}
with the unit norm
\begin{equation}
<l|l> = 1.
\end{equation}
Then, from the relation
\begin{equation}
\hat{\vec{L}}^2 = \hat{L}_- \hat{L}_{+} + \hat{L}_z (1+ \hat{L}_z),
\end{equation}
we obtain
\begin{equation}
\hat{\vec{L}}^2 |l> = l(l+1) |l>.
\end{equation}
By applying $\hat{L}_{-}$ to the highest-weight state $|l>$ several times, 
we obtain lower-lying states such as,
\begin{equation}
|l,m> \simeq (\hat{L}_-)^{l-m}|l>.
\end{equation}
Then we have the important results
\begin{eqnarray}
\hat{\vec{L}}^2 |l,m> &=& l(l+1) |l,m>,\\
\hat{L}_z |l,m> &=& m |l,m>,
\end{eqnarray}
where the first relation follows from 
$$ [\hat{\vec{L}}^2, \hat{L}_{\pm}]=0.$$

Note that we obtain

\begin{equation}
\hat{L}_- |l,-l>=0  
\end{equation}
since $||\hat{L}_-|l,-l>||^2 = 0.$ This equality follows from
$$ [\hat{L}_+, \hat{L}_-^{2l+1}]|l>=0.$$

Then we obtain
\begin{equation}
-l \leq m \leq +l. \label{eq:lm}
\end{equation}

We finally write down the explicit form of the normalized states as
\begin{equation}
|l,m> = \sqrt{\frac{(l+m)!}{(l-m)! (2l)!}} (\hat{L}_-)^{l-m}|l>.
\end{equation}
Another representation is 
\begin{equation}
|l,\pm|m|> = C(l,|m|)  (\hat{L}_{\pm})^{|m|} |l,0>,
\end{equation}
where ~ $ C(l,|m|) \equiv \sqrt{(l-|m|)!/(l+|m|)!}.$
\\

The usual wave function of the angle (SHs) can be given 
by specifying the representation such as $(\theta , \phi)$ gives;

\begin{equation}
Y_{lm} = <\theta, \phi|l,m>.
\end{equation}

The ``bra'' $<\theta, \phi|$ denotes the representation and the 
``ket" $|l,m>$ denotes the state. 
The inner product of these two vectors gives the usual wave function.

\section{Directional Parity (Mirror) Operator}
We define the space inversion (mirror) operator for each direction, 
$\hat{P}_x,~\hat{P}_y,~\hat{P}_z$ 
in the following \cite{Bjorken}:
\begin{eqnarray}
\hat{P}_x(x,p_x, other)\hat{P}_x^{-1} &=& (-x,-p_x,  other),\\
\hat{P}_y(y,p_y, other)\hat{P}_y^{-1} &=& (-y,-p_y,  other),\\
\hat{P}_z(z,p_z, other)\hat{P}_z^{-1} &=& (-z,-p_z,  other).
\end{eqnarray}

Therefore, from eqs.(\ref{eq:Lx}), (\ref{eq:Ly}), and (\ref{eq:Lz}), we obtain
\begin{eqnarray}
\hat{P}_x \hat{L}_x \hat{P}_x^{-1} &=&  \hat{L}_x,~ \hat{P}_y \hat{L}_x \hat{P}_y^{-1} = -\hat{L}_x,\nonumber \\
\hat{P}_z \hat{L}_x \hat{P}_z^{-1} &=& -\hat{L}_x,\\
\hat{P}_x \hat{L}_y \hat{P}_x^{-1} &=& -\hat{L}_y,~ \hat{P}_y \hat{L}_y \hat{P}_y^{-1} =  \hat{L}_y,\nonumber \\
\hat{P}_z \hat{L}_y \hat{P}_z^{-1} &=& -\hat{L}_y,\\
\hat{P}_x \hat{L}_z \hat{P}_x^{-1} &=& -\hat{L}_z,~ \hat{P}_y \hat{L}_z \hat{P}_y^{-1} = -\hat{L}_z,\nonumber \\
\hat{P}_z \hat{L}_z \hat{P}_z^{-1} &=&  \hat{L}_z.
\end{eqnarray}

Then the relation $[\hat{\vec{L}}^2, \hat{P}_k] =0$ follows, as is expected.
Furthermore, we have the trivial condition

\begin{equation}
\hat{P}_x^2 = \hat{P}_y^2 = \hat{P}_z^2 =1.
\end{equation}

The product of two different mirror operators is a rotation operator, for example,
\begin{eqnarray}
\hat{P}_x \hat{P}_y = e^{i\hat{L}_z \pi},\\
\hat{P}_y \hat{P}_z = e^{i\hat{L}_x \pi},\\
\hat{P}_z \hat{P}_x = e^{i\hat{L}_y \pi},
\end{eqnarray}

from which the following interesting property is obtained:
\begin{eqnarray}
\hat{P}_x \hat{L}_z |l,m> &=& \hat{P}_x \hat{L}_z \hat{P}_x^{-1}(\hat{P}_x |l,m>)
= - \hat{L}_z (\hat{P}_x |l,m>)\nonumber\\
&=& m (\hat{P}_x |l,m>).\nonumber
\end{eqnarray}

A similar relation also holds for $\hat{P}_y$. Therefore, we have

\begin{eqnarray}
\hat{L}_z (\hat{P}_x |l,m>) &=& -m (\hat{P}_x |l,m>),\\
\hat{L}_z (\hat{P}_y |l,m>) &=& -m (\hat{P}_y |l,m>).
\end{eqnarray}
However for $\hat{P}_z$, we have
\begin{equation}
\hat{L}_z (\hat{P}_z |l,m>) = +m (\hat{P}_z |l,m>).
\end{equation}

Thus, we can assume the following three equations:

\begin{eqnarray}
\hat{P}_x |l,m> &=& \alpha_x(l,m) |l,-m>,\\
\hat{P}_y |l,m> &=& \alpha_y(l,m) |l,-m>,\\
\hat{P}_z |l,m> &=& \alpha_z(l,m) |l,m>,
\end{eqnarray}
where $\alpha_k(l,m)$ is an unknown c-number.

The $m=0$ state has rotational symmetry around the $z$-axis since
$$ L_z |l,0>=0,~~ \mbox{and} ~~e^{-i \hat{L}_z \phi}|l,0>= |l,0>.$$

Thus, the state should have $x$-axis and $y$-axis mirror symmetry:
\begin{equation}
\hat{P}_x |l,0> = |l,0>,~~ \hat{P}_y |l,0> = |l,0>.
\end{equation}

Then we obtain 
\begin{eqnarray}
\hat{P}_x|l,\pm|m|> &=& C(l,|m|) \hat{P}_x(\hat{L}_{\pm})^{|m|}\hat{P}_x^{-1}|l,0>,\nonumber \\
&=&  C(l,|m|)  (\hat{L}_{\mp})^{|m|} |l,0>,\nonumber \\
&=& |l, \mp|m|>,
\end{eqnarray}
where we utilized 
$$\hat{P}_x \hat{L}_{\pm} \hat{P}_x^{-1} = \hat{L}_{\mp}.$$
On the other hand,
\begin{eqnarray}
\hat{P}_y|l,\pm|m|> &=& C(l,|m|) \hat{P}_y(\hat{L}_{\pm})^{|m|}\hat{P}_y^{-1}|l,0>,\nonumber \\
&=&  C(l,|m|)  (-1)^{|m|} (\hat{L}_{\mp})^{|m|} |l,0>,\nonumber \\
&=& (-1)^{|m|}|l, \mp|m|>,
\end{eqnarray}

where we utilized the relation
$$\hat{P}_y \hat{L}_{\pm} \hat{P}_y^{-1} = -\hat{L}_{\mp}.$$

Therefore, we obtain
\begin{equation}
\alpha_x =1, ~~~ \alpha_y = (-1)^m.
\end{equation}

We obtain $\alpha_z$ as follows.
From $\hat{P}_x \hat{P}_z = e^{i\pi \hat{L}_y}$ and an explicit form of 
$e^{i\pi \hat{L}_y}$ obtained from the $SU(2)$ representation 
(See (\ref{eq:y-rot1}) for $l=1$ and (\ref{eq:y_rot2}) for $l=2$), we have 
$$e^{i\pi \hat{L}_y}|l,m>=(-1)^{l+m}\hat{P}_x|l,m>.$$ 

Thus, we obtain $\alpha_z(l,m) = (-1)^{l+m}$.

To summarize,
\begin{eqnarray}
\hat{P}_x |l,m> &=&  |l,-m>,\label{eq:alphax}\\
\hat{P}_y |l,m> &=& (-1)^m |l,-m>,\label{eq:alphay}\\
\hat{P}_z |l,m> &=& (-1)^{l+m} |l,m>.\label{eq:alphaz}
\end{eqnarray}

The mirror operator in arbitrary direction is discussed in appendix.

\section{$l=0$ (s-state) case}
We start with the trivial case $l=0$.
We have
\begin{equation}
\hat{\vec{L}}^2 |s>=0
\end{equation}
for s-state $|s>$.
Then the eigen value of $\hat{L}_z$ should be zero from (\ref{eq:lm}). 
Therefore, we only have the $m=0$ state, which means that
\begin{equation}
\hat{L}_+ |s>=\hat{L}_-|s>=0.
\end{equation}
Then we obtain
\begin{equation}
\hat{L}_x |s>=\hat{L}_y|s>= \hat{L}_z|s>=0.
\end{equation}
These equations imply the following rotational invariance:
\begin{eqnarray}
e^{-i\hat{L}_x \theta_x} |s>&=&|s>,\\
e^{-i\hat{L}_y \theta_y} |s>&=&|s>,\\
e^{-i\hat{L}_z \theta_z} |s>&=&|s>,
\end{eqnarray}
where $\theta_x$, $\theta_y$, and $\theta_z$ are arbitrary independent angles.
Then the state $|s>$ should satisfy
\begin{equation}
Y_{00}(\theta,\phi) = <\theta, \phi|s> = const.
\end{equation}
Note that when we illustrate the form of the angle wave function,
we take the radial length $r=|Y_{lm}(\theta,\phi)|$ as the magnitude of wave function,
and show the wave function as the surface $r=r(\theta, \phi)$.
Thus, the form of wave function given by $r = |Y_{00}(\theta,\phi)|$ is a sphere.

\section{$l=1$ (p-state) case}
Let us start to find the form of the $l=1$ states.
First we define the state vectors,
\begin{equation}
|1,1> = \ \left(
    \begin{array}{c}
      1\\
      0\\
       0   \end{array}
  \right),
|1,0> = \ \left(
    \begin{array}{c}
      0\\
      1\\
       0   \end{array}
  \right),
|1,-1> = \ \left(
    \begin{array}{c}
      0\\
      0\\
      1   \end{array}
  \right).
\end{equation}
Then the representation of the angular momentum
\begin{equation}
[{\bf L_j}]_{mn} \equiv <1,m| \hat{L}_j |1,n>,~~j=1,2,3
\end{equation}
takes the following form (In the matrix representation,
 we write operators in bold):
\begin{eqnarray}
{\bf L_x} &=& \frac{1}{\sqrt{2}}\ \left(
    \begin{array}{ccc}
      0& 1 & 0 \\
      1 & 0 & 1\\
      0 & 1 & 0    \end{array}
  \right),~~
{\bf L_y} = \frac{i}{\sqrt{2}}\ \left(
    \begin{array}{ccc}
      0& -1 & 0 \\
      1 & 0 & -1\\
      0 & 1 & 0    \end{array}
  \right),\nonumber\\
&& ~~~~~~~~~{\bf L_z} = \ \left(
    \begin{array}{ccc}
      1& 0 & 0 \\
      0& 0 & 0\\
      0 & 0& -1    \end{array}
  \right).
\end{eqnarray}

Then the rotational matrix can be calculated by the 
Taylor expansion of the following matrix-valued exponent:
$$ e^{-i {\bold L} \phi} = \sum_{n=0}^\infty \frac{1}{n!} (-i {\bold L} \phi)^n.$$

To carry out this calculation, we predict the general term $(-i {\bold L} \phi)^n$ and prove it by mathematical induction. Then we calculate the sum of the series.
We obtain

\begin{eqnarray}
e^{-i{\bf L_x} \phi}&=& \frac{1}{2} \ \left(
    \begin{array}{ccc}
      1& 0& -1 \\
      0 & 0 & 0\\
      -1 & 0 & 1   \end{array}
  \right) 
+  \frac{1}{2} \ \left(
    \begin{array}{ccc}
      1& 0 & 1 \\
      0& 2 & 0\\
      1 & 0 & 1    \end{array}
  \right) \cos \phi \nonumber\\
&&  -  \frac{i}{\sqrt{2}}\ \left(
    \begin{array}{ccc}
      0& 1 & 0 \\
      1 & 0 & 1\\
      0 & 1 & 0    \end{array}
  \right) \sin \phi, \label{eq:x-rot1}
  \end{eqnarray}

\begin{eqnarray}
e^{-i{\bf L_y} \phi}&=& \frac{1}{2} \ \left(
    \begin{array}{ccc}
      1& 0& 1 \\
      0 & 0 & 0\\
      1 & 0 & 1   \end{array}
  \right) 
+  \frac{1}{2} \ \left(
    \begin{array}{ccc}
      1& 0 & -1 \\
      0& 2 & 0\\
     - 1 & 0 & 1    \end{array}
  \right) \cos \phi \nonumber\\
&&  -  \frac{1}{\sqrt{2}}\ \left(
    \begin{array}{ccc}
      0& 1 & 0 \\
      -1 & 0 & 1\\
      0 & -1 & 0    \end{array}
  \right) \sin \phi,\label{eq:y-rot1}
\end{eqnarray}

\begin{eqnarray}
e^{-i{\bf L_z} \phi}&=& \ \left(
    \begin{array}{ccc}
      0& 0& 0 \\
      0 & 1 & 0\\
      0& 0 & 0  \end{array}
  \right) 
+   \ \left(
    \begin{array}{ccc}
      1& 0 & 0 \\
      0& 0 & 0\\
      0 & 0 & 1    \end{array}
  \right) \cos \phi \nonumber\\
&&  -  i\ \left(
    \begin{array}{ccc}
      1& 0& 0 \\
      0 & 0 & 0\\
      0 & 0& -1   \end{array}
  \right) \sin \phi.\label{eq:z-rot1}
\end{eqnarray}

Next we construct the real spherical harmonics (hereafter abbreviated to RSHs)
 to show their form graphically.
The SH itself is a complex function, 
and the method of constructing the RSH from the SH is well known \cite{Chisholm}.
In accordance with this construction, we define the following states:

\begin{eqnarray}
|x> &\equiv& \frac{1}{\sqrt{2}} \ \left(
    \begin{array}{c}
      -1\\
      0\\
      1   \end{array}
  \right)
= \frac{1}{\sqrt{2}} (|1,-1> -|1,1>), \label{eq:state_x}\\
|y> &\equiv& \frac{i}{\sqrt{2}} \ \left(
    \begin{array}{c}
      1\\
      0\\
      1   \end{array}
  \right)
= \frac{i}{\sqrt{2}} (|1,1> +|1,-1>),\label{eq:state_y}\\
|z> &\equiv& \ \left(
    \begin{array}{c}
      0\\
      1\\
      0   \end{array}
  \right)
= |1,0> .\label{eq:state_z}
\end{eqnarray}

\begin{figure}[h]
\begin{center}
\includegraphics[width=2.5cm]{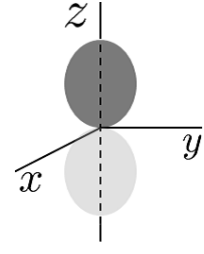}
\end{center}
\caption{Schematic diagram of $|z>$ state}
\end{figure}

Next we discuss the form of the states.
We start with the $|z>$ state.
First, this state has rotational symmetry around the $z$-axis:
\begin{equation}
e^{-i{\bf L_z} \phi}|z> = |z>.
\end{equation}
Second, the $|z>$ state has parity odd for $\hat{P}_z$ from (\ref{eq:alphaz}), 
which means that the state has $xy$-plane as the node plane.
\begin{equation}
{\bf P_z}|z> = -|z>.
\end{equation}

Then the form of $|z>$ ($r=|<\theta,\phi|z>|$) is considered to be that in figure 1 , 
where the different contrasts show the phase inversion.
Figure 1 is a rough sketch and the precise form will be discussed later.
Next, we consider the forms of the other two states.

First, from (\ref{eq:x-rot1}), (\ref{eq:state_y}), and (\ref{eq:state_z}), we obtain

\begin{equation}
e^{-i{\bf L_x} (-\pi/2)} |z> = |y>. \label{eq:z-y}
\end{equation}

Second, from (\ref{eq:y-rot1}), (\ref{eq:state_x}), and (\ref{eq:state_z}), we obtain

\begin{equation}
e^{-i{\bf L_y} (\pi/2)} |z> = |x>.\label{eq:z-x}
\end{equation}

Therefore, these three states have the same form but different orientations.
We obtain the form of $|z>$ more precisely as follows.
The rotation of the $|z>$ state around the $y$ axis by angle $\alpha$ gives the following relation from 
 (\ref{eq:y-rot1}), (\ref{eq:state_x}), and (\ref{eq:state_z}):

\begin{figure}[h]
\begin{center}
\includegraphics[width=9cm]{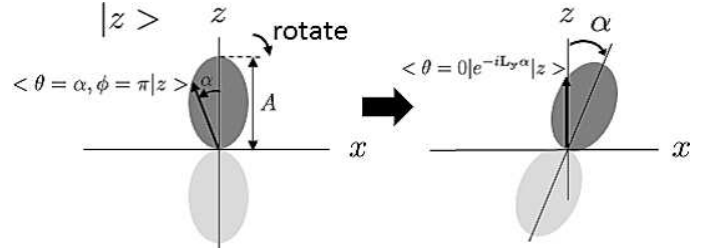}
\end{center}
\caption{Rotation of $|z>$ around $y$ axis by angle $\alpha$}
\end{figure}

\begin{equation}
e^{-i{\bf L_y} \alpha} |z> = \cos \alpha |z> + \sin \alpha |x>.
\end{equation}

Then we examine the $z$ direction. For this purpose, we multiply "bra" $<\theta=0|$ from the left. 
We obtain
\begin{eqnarray}
<\theta=0|e^{-i{\bf L_y} \alpha} |z> &=& \cos \alpha <\theta=0|z> \nonumber\\
&+& \sin \alpha <\theta=0|x>.\label{eq:l=1_rot_exp}
\end{eqnarray}

The l.h.s. can be calculated as
$$e^{+i{\bf L_y} \alpha}|\theta=0>= |\theta=\alpha, \phi=\pi>.$$
By taking the Hermitian conjugate, we obtain
\begin{eqnarray}
<\theta=0|e^{-i{\bf L_y} \alpha} |z> &=& <\theta=\alpha, \phi=\pi|z> \nonumber\\
&=& <\theta=\alpha|z>,\label{eq:l=1_rot}
\end{eqnarray}
where the final equality originates from the rotational symmetry of $|z>$ about the $z$ axis.
This situation is shown in figure 2.
$<\theta=\alpha,\phi=\pi|z>$ is shown by the arrow in the left figure.
To obtain the length of this arrow, we rotate the state $|z>$ 
around the $y$ axis by angle $\alpha$ and examine the $z$ direction.

Furthermore, from the form of $|x>$, we have 
\begin{equation}
<\theta=0|x>=0. \label{eq:x0}
\end{equation}
This comes from (\ref{eq:z-x}) and figure 1, later explicitly shown in figure 4.
From (\ref{eq:l=1_rot_exp}), (\ref{eq:l=1_rot}), and (\ref{eq:x0}), we have
\begin{equation}
<\theta=\alpha|z> = A \cos \alpha,
\end{equation}
where the constant $A$ is given by $A \equiv <\theta=0|z>$ 
and we set $A$ to be real and positive.

Then the wave function can be written as
\begin{equation}
Y_{1,z}(\theta, \phi) \equiv <\theta, \phi|z>= A \cos \theta. \label{eq:1z_wave_function}
\end{equation}
To illustrate this, let $r= |Y_{1,z}(\theta,\phi)|$. Then
$$ r = A \frac{z}{r} ~~(z>0),~~  r = -A \frac{z}{r} ~~(z<0).$$
Then we have
\begin{equation}
x^2 + y^2 + (z \pm A/2)^2 = (A/2)^2, ~~~ \mbox{for} ~z \lessgtr 0. \label{eq:1z_form}
\end{equation}

From (\ref{eq:1z_wave_function}), the two spheres with centers $(0,0,A/2)$ 
and $(0,0,-A/2)$ have opposite phases. Thus, we show them 
with different contrast in figure 3.

\begin{figure}[h]
\begin{center}
\includegraphics[width=3.5cm]{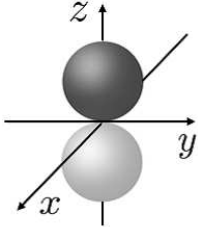}
\end{center}
\caption{Precise form of $|z>$}
\end{figure}

To conclude this section, we show all the  forms of the 
$l=1$ members i.e., the $|x>, |y>,$ and $|z>$ states, and their relations 
in figure 4, as obtained from (\ref{eq:z-y}), (\ref{eq:z-x}),
 and (\ref{eq:1z_form}).

\begin{figure}[h]
\begin{center}
\includegraphics[width=8cm]{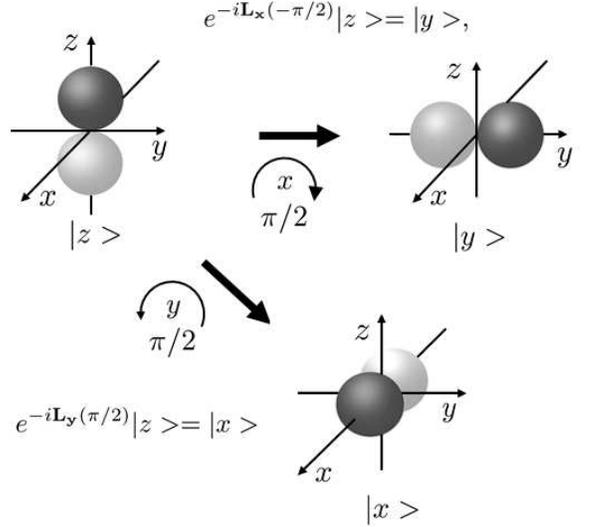}
\end{center}
\caption{Forms of the $l=1$ members}
\end{figure}

\section{$l=2$ (d-state) case}
Next we consider the $l=2$ case.
The matrix elements of the angular moment can be calculated as
\begin{equation}
[{\bf L_j}]_{mn} \equiv <2,m| \hat{L}_j |2,n>,  ~~~j=1,2,3,
\end{equation}
with the notation

\begin{eqnarray}
|2,2> &=& \ \left(
\begin{array}{c}
1\\
0\\
0\\
0\\
0 \end{array}
  \right),
~
|2,1> = \ \left(
    \begin{array}{c}
      0\\
      1\\
      0\\
      0\\
      0   
\end{array}
\right),
~
|2,0> = \ \left(
    \begin{array}{c}
      0\\
      0\\
      1\\
      0\\
      0   
\end{array}
\right),\nonumber\\
|2,-1> &=& \ \left(
    \begin{array}{c}
      0\\
      0\\
      0\\
      1\\
      0   
\end{array}
\right),
~
|2,-2> = \ \left(
    \begin{array}{c}
      0\\
      0\\
      0\\
      0\\
      1   
\end{array}
\right).
\end{eqnarray}

Then we have the explicit matrix forms
\begin{eqnarray}
{\bf L_x} &=&\ \left(
    \begin{array}{ccccc}
      0& 1 & 0 &0&0\\
      1 & 0 & \sqrt{6}/2& 0& 0\\
      0 &\sqrt{6}/2 & 0 & \sqrt{6}/2 & 0\\
      0& 0 & \sqrt{6}/2 & 0&1\\
      0&0&0&1&0 
\end{array}
\right),\\
{\bf L_y} &=&i\ \left(
    \begin{array}{ccccc}
      0& -1 & 0 & 0&0\\
      1 & 0 & -\sqrt{6}/2& 0&0\\
      0 & \sqrt{6}/2&0&-\sqrt{6}/2&0\\
      0&0&\sqrt{6}/2&0&-1\\
      0&0&0&1&0    
\end{array}
\right),\\
{\bf L_z} &=& \ \left(
    \begin{array}{ccccc}
      2 & 0 & 0 & 0 & 0\\
      0 & 1 & 0 & 0 & 0\\
      0 & 0 &  0& 0 &0\\
      0 & 0 & 0 & -1 & 0\\
      0 & 0 & 0 & 0 & -2\\
\end{array}
\right).
\end{eqnarray}

Then the rotation matrices around 
the $x$, $y$, and $z$ axes are calculated by the Taylor expansion
 of the matrix-valued exponents,
the same way as the case of $l=1$: 

\begin{eqnarray}
e^{-i {\bf L_x} \phi}  &=&\ \left(
    \begin{array}{ccccc}
      A& iB & C &iD&E\\
      iB & F & iG& H& iD\\
      C &iG& J& iG & C\\
      iD& H & iG & F&iB\\
     E&iD&C&iB&A \end{array}
  \right),\label{eq:x_rot2}\\
e^{-i {\bf L_y} \phi}  &=&\ \left(
    \begin{array}{ccccc}
      A& B & -C &-D&E\\
      -B & F & G& -H& -D\\
      -C &-G& J& G & -C\\
      D& -H & -G & F&B\\
     E&D&-C&-B&A \end{array}
  \right),\label{eq:y_rot2}\\
e^{-i {\bf L_z} \phi}  &=&\ \left(
    \begin{array}{ccccc}
      e^{-2i\phi}& 0 & 0 &0&0\\
      0 & e^{-i\phi} & 0& 0& 0\\
      0 &0& 1& 0 & 0\\
      0& 0 & 0&e^{+i\phi} &0\\
     0&0&0&0&e^{+2i\phi} \end{array}
  \right),\label{eq:z_rot2}
  \end{eqnarray}
where
\begin{eqnarray}
A &=& \frac{3}{8} + \frac{1}{8} \cos 2\phi + \frac{1}{2} \cos \phi,\nonumber\\
B &=& -\frac{1}{2} \sin \phi - \frac{1}{4} \sin 2\phi,\nonumber\\
C &=& \frac{\sqrt{6}}{8} (\cos 2\phi -1)\nonumber\\
D &=& -\frac{1}{4} \sin 2\phi + \frac{1}{2}\sin \phi,\nonumber\\
E &=& \frac{3}{8} + \frac{1}{8} \cos 2\phi -\frac{1}{2} \cos \phi,\nonumber\\
F &=& \frac{1}{2} (\cos \phi + \cos 2\phi),\nonumber\\
G &=& -\frac{\sqrt{6}}{4} \sin 2\phi,\nonumber\\
H &=& \frac{1}{2} (\cos 2\phi - \cos \phi),\nonumber\\
J &=& \frac{1}{4} + \frac{3}{4} \cos 2\phi.
\end{eqnarray}
The RSHs are given as follows \cite{Chisholm}:
\begin{eqnarray}
|xy> &\equiv&  -\frac{i}{\sqrt{2}}(|2,2> - |2,-2>),\nonumber\\
|x^2-y^2> &\equiv& \frac{1}{\sqrt{2}}(|2,2> + |2,-2>),\nonumber\\
|yz> &\equiv&  \frac{i}{\sqrt{2}}(|2,1> + |2,-1>), \nonumber\\
|xz> &\equiv&  -\frac{1}{\sqrt{2}}(|2,1> - |2,-1>),\nonumber\\
|z^2> &\equiv& |2,0>.
\end{eqnarray}

We start with the analysis of $|xy>$.
The first observation is the rotation of $|xy>$ around the $z$ axis by $-\pi/2$:
\begin{equation}
e^{i {\bf L_z} (\pi/2)} |xy> =  -  |xy>.
\end{equation}
This means that the state is fourfold symmetric except for the phase change $e^{i\pi}$.
Second, the state has two node planes,
\begin{eqnarray}
{\bf P_x}|xy> &=&-\frac{i}{\sqrt{2}} {\bf P_x} (|2,2> - |2,-2>)\nonumber\\
&=& -\frac{i}{\sqrt{2}}  (|2,-2> - |2,2>)=-|xy>,\\
{\bf P_y}|xy> &=& -\frac{i}{\sqrt{2}} {\bf P_y} (|2,2> - |2,-2>)\nonumber\\
&=& -\frac{i}{\sqrt{2}} (|2,-2> - |2,2>)=-|xy>.
\end{eqnarray}
This means that the $yz$ plane and the $xz$ plane are node planes.
Then we have the form of $|xy>$ shown in figure 5.

\begin{figure}[h]
\begin{center}
\includegraphics[width=4cm]{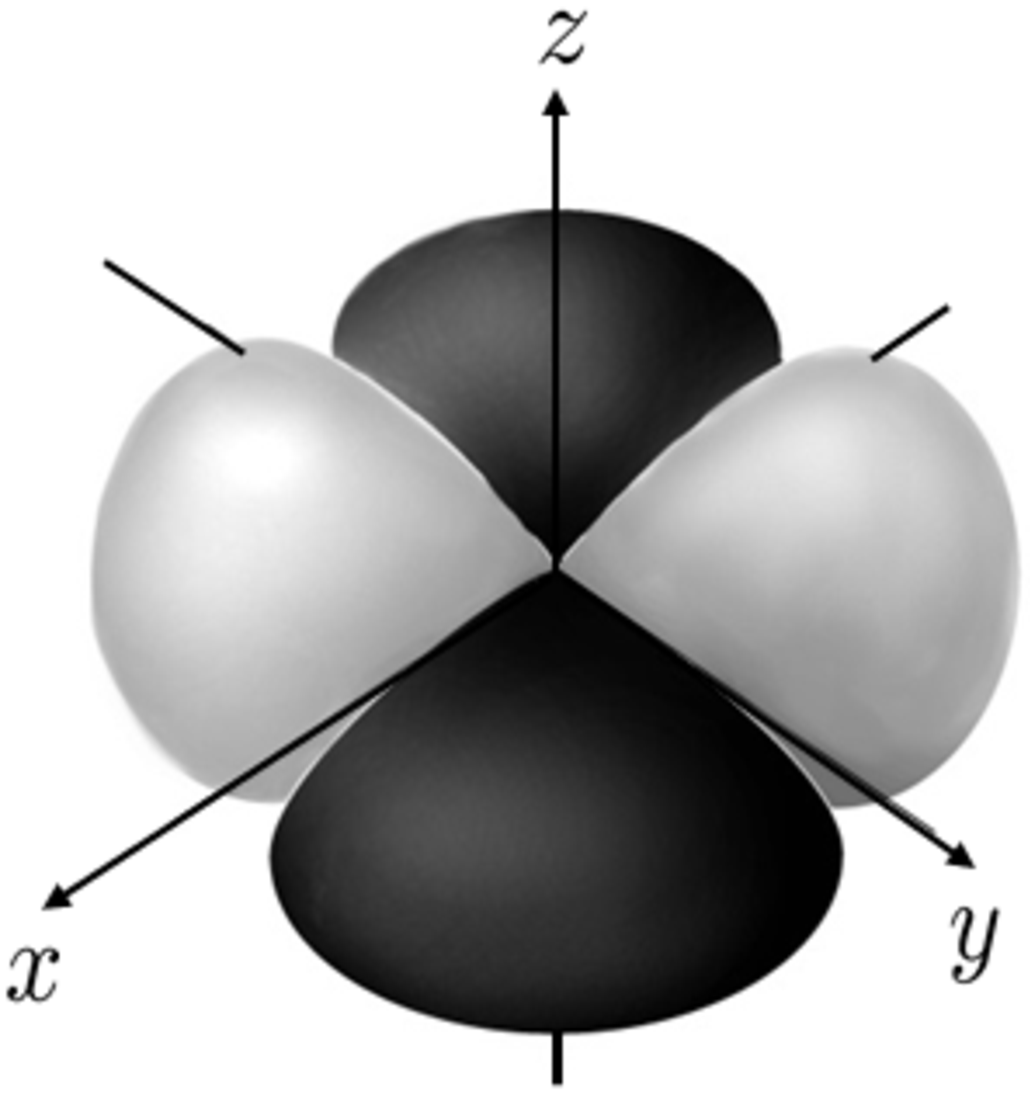}
\end{center}
\caption{Form of $|xy>$}
\end{figure}

Another three states can be constructed easily from $|xy>$ as follows:

\begin{eqnarray}
e^{i {\bf L_y} (\pi/2)} |xy> &=& \frac{i}{\sqrt{2}}
\ \left(
    \begin{array}{c}
      0\\
      1 \\
      0\\
      1\\
     0\end{array}
  \right)
=|yz>,\label{eq:fig(yz)}\\
e^{-i {\bf L_x} (\pi/2)} |xy> &=& -\frac{1}{\sqrt{2}}
\ \left(
    \begin{array}{c}
      0\\
      1 \\
      0\\
      -1\\
     0\end{array}
  \right)
= |xz>,\\
e^{i {\bf L_z} (\pi/4)} |xy> &=& \frac{1}{\sqrt{2}}
\ \left(
    \begin{array}{c}
      1\\
      0 \\
      0\\
      0\\
     1\end{array}
  \right)
=|x^2-y^2>.\label{eq:fig(x^2-y^2)}
\end{eqnarray}
Therefore, these $|yz>$, $|xz>$, and $|x^2-y^2>$ states have the same form as $|xy>$ 
but different orientations.

\section{$|z^2>$ state}
Finally, we consider the form of $|z^2>$.
We have two symmetries,
\begin{eqnarray}
e^{-i{\bf L_z} \phi}|z^2> &=& |z^2>,\label{eq:axial_z}\\
{\bf P_z}|z^2> &=& |z^2>.
\end{eqnarray}
These equations show rotational symmetry around the $z$ axis, 
and reflection (mirror) symmetry about the $xy$ plane,
that are insufficient informations to construct the form of the state $|z^2>$.

Let us rotate $|z^2>$ around the $x$ axis by angle $-\alpha$.
\begin{eqnarray}
&&e^{i {\bf L_x} \alpha}|z^2> = \ \left(
    \begin{array}{ccccc}
      A& iB & C &iD&E\\
      iB & F & iG& H& iD\\
      C &iG& J& iG & C\\
      iD& H & iG & F&iB\\
     E&iD&C&iB&A \end{array}
  \right)
\ \left(
 \begin{array}{c}
      0\\
      0\\
     1\\
      0\\
    0 \end{array}
  \right)  \nonumber\\
&&=
\ \left(
\begin{array}{c}
      C\\
      iG\\
     J\\
      iG\\
    C\end{array} \right) =  \sqrt{2} C(-\alpha) |x^2-y^2> \nonumber\\
&& ~~~~~~~~~~~~~~~~ + \sqrt{2} G(-\alpha) |yz> + J(-\alpha) |z^2>. \label{eq:expand_z^2}
\end{eqnarray}
Let us examine the $z$ direction. 
We multiply the ``bra" $<\theta=0|$ to both sides of eq. (\ref{eq:expand_z^2}).
\begin{eqnarray}
&&<\theta=0|e^{i {\bf L_x} \alpha} |z^2> =  \sqrt{2} C(-\alpha)<\theta=0 |x^2-y^2>\nonumber \\
&&+ \sqrt{2} G(-\alpha) <\theta=0|yz> \nonumber\\
&&+ J(-\alpha) <\theta=0|z^2>.
\end{eqnarray}
Note that $<\theta=0 |x^2-y^2>=<\theta=0|yz>=0$ hold here.
This comes from the following reasons.
From (\ref{eq:fig(x^2-y^2)}) and Figure 5, we have $<\theta=0 |x^2-y^2>=0$.
From (\ref{eq:fig(yz)}) and Figure 5, we have $<\theta=0|yz>=0$. 
Both are later shown in figure 9 explicitly.
Furthermore, from
$$e^{-i {\bf L_x}\alpha}|\theta=0> = |\theta=\alpha, \phi=-\pi/2>,$$
we obtain
\begin{eqnarray}
<\theta=0|e^{i {\bf L_x} \alpha} |z^2> &=& <\theta=\alpha, \phi=-\pi/2|z^2>\nonumber\\
&=& <\theta=\alpha|z^2>,
\end{eqnarray}
where the final equality originates from the rotational symmetry around the $z$ axis from (\ref{eq:axial_z}).
Then we obtain
\begin{equation}
<\theta=\alpha|z^2> = J(-\alpha) <\theta=0|z^2>.
\end{equation}
In an explicit form, we have
\begin{equation}
Y_{z^2}(\theta,\phi) = \frac{l_0}{4} (1+3\cos 2\theta),~~~ l_0 \equiv Y_{z^2}(\theta=0). \label{eq:y_z^2}
\end{equation}

\begin{figure}[h]
\begin{center}
\includegraphics[width=5cm]{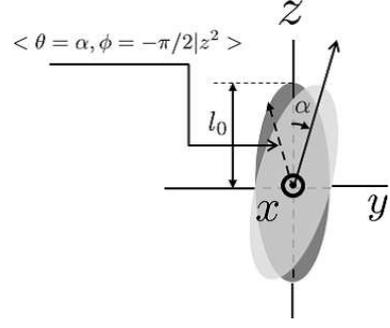}
\end{center}
\caption{Method of obtaining the form of $|z^2>$}
\end{figure}

This method is graphically shown in figure 6.
The dark gray ellipsoid shows the $|z^2>$ state 
and the light gray ellipsoid shows the state $|z'^2> \equiv e^{i {\bf L_x} \alpha} |z^2>$.  
Then, we easily find that
$$<\theta=\alpha, \phi=-\pi/2|z^2>=<\theta=0|z'^2>.$$
(The length of the dashed arrow of $|z^2>$ is the same as the length of $|z'^2>$ in the $z$ direction.) Furthermore, $|z'^2>$ can be expanded in the form of eq. (\ref{eq:expand_z^2}).
We therefore obtain equation (\ref{eq:y_z^2}).
From this result, we have the form of $<\theta|z^2>$ shown in figure 7.

\begin{figure}[h]
\begin{center}
\includegraphics[width=4cm]{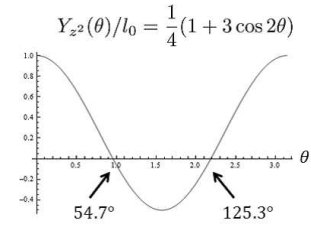}
\end{center}
\caption{Functional form of $<\theta|z^2>$}
\end{figure}

\begin{figure}[h]
\begin{center}
\includegraphics[width=3.5cm]{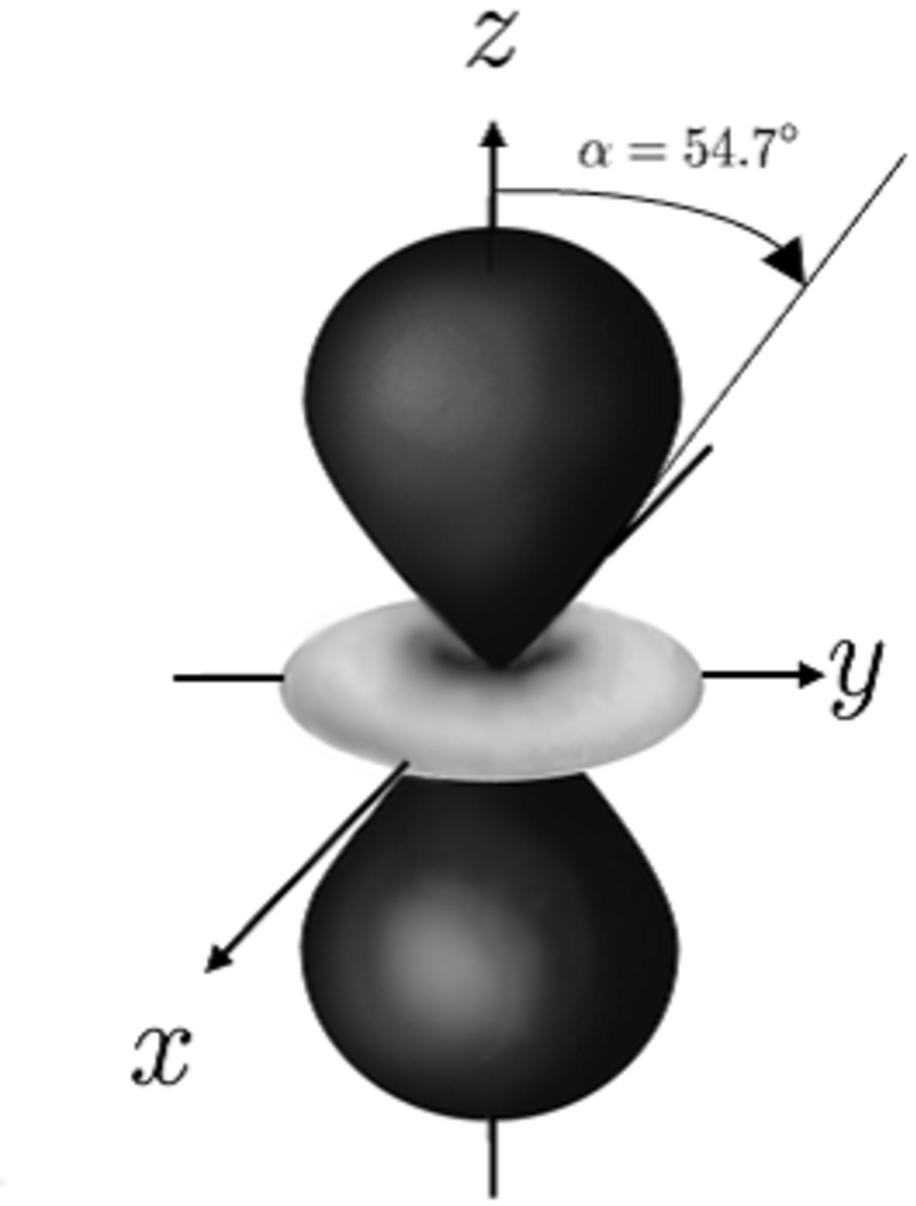}
\end{center}
\caption{Form of $<\theta|z^2>$}
\end{figure}

In figure 8, the dark gray part and light gray part (similar to a torus but with a point hole) 
have opposite phases. The node plane becomes two cones with 
$\theta=54.7^\circ$ and $\theta=125.3^\circ$.

\begin{figure}[h]
\begin{center}
\includegraphics[width=8cm]{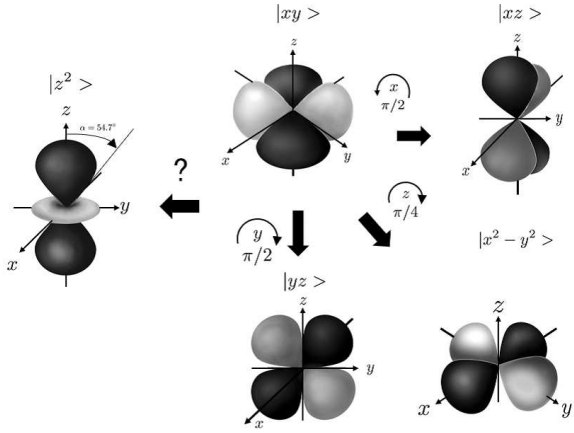}
\end{center}
\caption{Forms of the members of $l=2$}
\end{figure}

The states comprising the members of $l=2$ are shown in figure 9.
One of the remaining problems is the relation between $|z^2>$ and the other states.
From figure 9, we search for the states that may become elements to construct the $|z^2>$ state.
The rotation of $|xz>$ around the $y$ axis by $-\pi/4$ with the rotation of $|yz>$ around the $x$ axis by $\pi/4$
may have similar forms to $|z^2>$ as shown in figure 10.

\begin{figure}[h]
\begin{center}
\includegraphics[width=6cm]{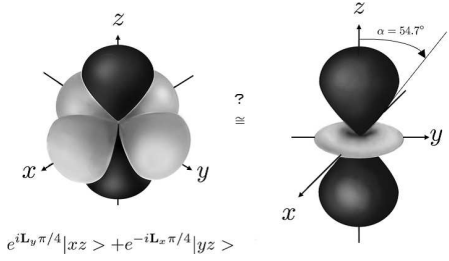}
\end{center}
\caption{Similar forms of $|z^2>$}
\end{figure}

This idea can be realized in the following calculation:

\begin{eqnarray}
&& e^{i{\bf L_y} \pi/4} |xz>  +  e^{-i{\bf L_x} \pi/4}|yz> \nonumber\\
&=& (-\frac{1}{2}|x^2-y^2> + \frac{\sqrt{3}}{2} |z^2>) \nonumber\\
&& +(\frac{1}{2}|x^2-y^2> + \frac{\sqrt{3}}{2} |z^2>)\nonumber\\
&=& \sqrt{3} |z^2>.\nonumber
\end{eqnarray}

Or alternatively,

\begin{equation}
|z^2> = \frac{1}{\sqrt{3}} (e^{i{\bf L_y} \pi/4} |xz>  +  e^{-i{\bf L_x} \pi/4}|yz>).
\end{equation}
In this way, we obtain the relation between $|z^2>$ and the other states.

\section{Functional form of $|xy>$ state}
To conclude the study of the $l=2$ state, we finally discuss the functional form of
the $|xy>$ state. From the fourfold property of the $|xy>$ state, 
 it is sufficient to study only one piece of four leaves of $|xy>$.
For this purpose, we focus on one leaf in the region $x>0,~ y>0$ of $|xy>$.
To obtain the wave function on the $xy$ plane with $\phi=\alpha$, we consider the wave function

 $$l(\alpha) = <\theta=\pi/2, \phi=\alpha|xy>.$$
 
To obtain $l(\alpha)$, we rotate the $|xy>$ state around the $z$ axis by angle $-\alpha$, 
and examine the $x$ direction as shown in figure 11.

\begin{figure}[h]
\begin{center}
\includegraphics[width=8cm]{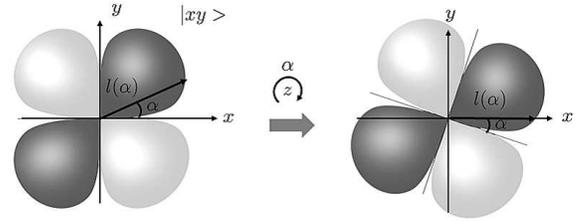}
\end{center}
\caption{Method used to obtain $l(\alpha)$}
\end{figure}

Using eq. (\ref{eq:z_rot2}), we obtain

\begin{equation}
e^{i {\bf L_z} \alpha} |xy> = \cos 2\alpha |xy> + \sin 2\alpha |x^2-y^2>.
\end{equation}

Then we have
\begin{eqnarray}
&& <\theta=\pi/2,\phi=0|e^{i {\bf L_z} \alpha}|xy> \nonumber\\
&=& \cos 2\alpha<\theta=\pi/2,\phi=0 |xy> \nonumber\\
&&~+ \sin 2\alpha<\theta=\pi/2,\phi=0 |x^2-y^2>.
\end{eqnarray}

The l.h.s. can be calculated as

\begin{equation}
e^{-i {\bf L_z} \alpha} |\theta=\pi/2,\phi=0> =  |\theta=\pi/2,\phi=\alpha>.
\end{equation}

The Hermitian conjugation gives

\begin{equation}
<\theta=\pi/2,\phi=0|e^{i {\bf L_z} \alpha}~  = ~ <\theta=\pi/2,\phi=\alpha|.
\end{equation}

Furthermore, by using

$$<\theta=\pi/2,\phi=0 |xy> =0,$$

we obtain

\begin{eqnarray}
&&<\theta=\pi/2,\phi=\alpha|xy> \nonumber\\
&& =  \sin 2\alpha<\theta=\pi/2,\phi=0 |x^2-y^2>.
\end{eqnarray}
This is the same as
\begin{eqnarray}
l(\alpha) &\equiv& <\theta=\pi/2,\phi=\alpha|xy> = L_0 \sin 2\alpha,\nonumber \\
L_0 &\equiv& <\theta=\pi/2,\phi=0 |x^2-y^2>.
\end{eqnarray}

The same discussion can be generalized to a fixed $\theta$ 
(i.e., a cone surface with $\theta =$const.)
\begin{eqnarray}
l(\theta,\alpha) &=& <\theta,\phi=\alpha|xy> = L(\theta) \sin 2\alpha, \nonumber\\
L(\theta) &\equiv& <\theta,\phi=0|x^2-y^2>, \label{eq:l}
\end{eqnarray}
where the quantity $L(\theta)$ is shown in figure 12.\\

\begin{figure}[h]
\begin{center}
\includegraphics[width=9cm]{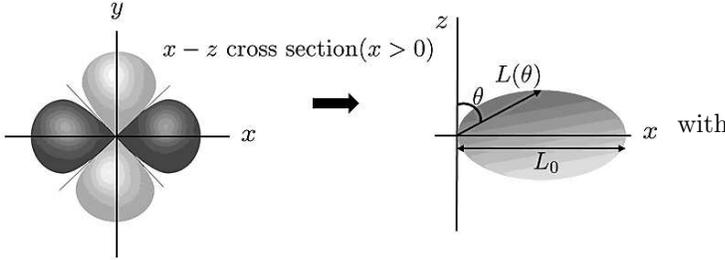}
\end{center}
\caption{Quantity $L(\theta)$ in state $|x^2-y^2>$}
\end{figure}

$L(\theta)$ is obtained by the method shown in figure 13.
To obtain $L(\alpha) = <\theta=\alpha, \phi=0|x^2-y^2>$, we rotate $|x^2-y^2>$ 
around $y$ axis by $-\alpha$, and examine the $z$ direction.
First, the rotation of state $|x^2-y^2>$ around the $y$ axis by $-\alpha$  is given by

\begin{figure}[h]
\begin{center}
\includegraphics[width=7cm]{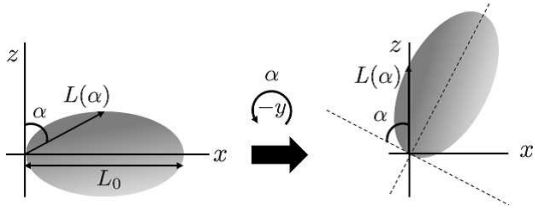}
\end{center}
\caption{Method to obtain $L(\theta)$}
\end{figure}

\begin{eqnarray}
&&e^{i {\bf L_y} \alpha} |x^2-y^2> = -\sqrt{2} C(-\alpha) |z^2>\nonumber\\
&& +(B(-\alpha)+D(-\alpha))|xz> \nonumber\\
&& + (A(-\alpha)+E(-\alpha))|x^2-y^2>.\label{eq:x^2-y^2}
\end{eqnarray}

Using
$$<\theta=0|xz>=<\theta=0|x^2-y^2>=0,$$
we obtain
\begin{eqnarray}
&& <\theta=0| e^{i {\bf L_y} \alpha} |x^2-y^2> \nonumber\\
&=& -\sqrt{2} C(-\alpha) <\theta=0|z^2> \nonumber\\
&=& \frac{\sqrt{3}}{4}(1-\cos 2 \alpha) <\theta=0|z^2>.
\end{eqnarray}
Furthermore from the relation
$$ e^{-i {\bf L_y} \alpha} |\theta=0> = |\theta=\alpha, \phi=0>,$$
we obtain
\begin{eqnarray}
L(\alpha)&=&<\theta=\alpha, \phi=0|x^2-y^2>\nonumber\\
&=& \frac{\sqrt{3}}{4}(1-\cos 2 \alpha) l_0, \label{eq:L}
\end{eqnarray}
where $l_0 \equiv <\theta=0|z^2>$.
From (\ref{eq:l}) and (\ref{eq:L}), we obtain
\begin{equation}
Y_{xy}(\theta,\phi)= \frac{\sqrt{3}}{4}l_0 (1-\cos 2 \theta) \sin 2\phi,
\end{equation}
with
$$L_0 = \frac{\sqrt{3}}{2} l_0.$$

\section{Conclusion}
We have shown a new method of obtaining spherical harmonics 
without solving the partial differential equation. 
This involves using the $SU(2)$ algebra and the directional space inversion (mirror) operator, 
where the latter was introduced in section 2.
The node plane is expressed in simple manner using this new operator.
Second, we have shown that the same $l$ states but different $m$ states are related to each other 
by the rotational group $SU(2)$. 
Solving the partial differential equation (\ref{eq:Laplacian}) 
is the simplest way to obtain the form of spherical harmonics; however, 
the physical relations between the solutions with different quantum numbers 
can also be understood using this method.

\section{acknowledgment}
The author would like to thank Ms. Kodera for her help in drawing the figures.

\section{Appendix-Constructing the Mirror Operator in an arbitrary direction}
We define the mirror operator in the $(\theta,\phi)$ direction as
\begin{equation}
\hat{P}(\theta, \phi).
\end{equation}

For example the mirror operators in the main text are expressed as
\begin{eqnarray}
\hat{P}_x &=& \hat{P}(\pi/2,0),\nonumber\\
\hat{P}_y &=& \hat{P}(\pi/2,\pi/2),\nonumber\\
\hat{P}_z &=& \hat{P}(0,\phi)~~ (\phi \mbox{~ is arbitrary}).\nonumber
\end{eqnarray}

The method of rotating $(\theta, \phi)$ in the $+x$ direction is as follows.
\begin{enumerate}
 \item $-\phi$ rotation around $z$ axis,
 \item $\pi/2-\theta$ rotation around $y$ axis.
\end{enumerate}

Therefore the space inversion into $(\theta, \phi)$ direction is given
by the following steps.

\begin{enumerate}
\item rotation around $z$ axis by $-\phi$,
\item rotation around $y$ axis by $\pi/2-\theta$,
\item space inversion in $x$ direction,
\item rotation around $y$ axis by $-\pi/2+\theta$,
\item rotation around $z$ axis by $+\phi$.
\end{enumerate}

Then we obtain the following general formula for the mirror operator 
in an arbitrary direction:

\begin{equation}
\hat{P}(\theta, \phi)=e^{-i\hat{L}_z\phi} e^{i\hat{L}_y(\pi/2-\theta)} 
\hat{P}_x e^{-i\hat{L}_y(\pi/2-\theta)} e^{i\hat{L}_z\phi} .
\end{equation}

Using the matrix form of $\hat{P}_x$,

\begin{equation}
(P_x)_{m,n} = \delta_{m,-n},
\end{equation}

we can directly verify the relation by using the general formula
\begin{eqnarray}
(P_y)_{m,n} &=& (-1)^m\delta_{m,-n},\\
(P_z)_{m,n} &=& (-1)^{m+l}\delta_{m,n}.
\end{eqnarray}

\end{document}